\documentclass[hyper]{JHEP3}

\input{epsf}
\usepackage{epsfig}
\usepackage{amssymb}
\usepackage{amsfonts}
\usepackage{amsbsy}
\usepackage[all,2cell]{xy}

\usepackage{amsmath}

\usepackage{amssymb,amscd}
\usepackage{mathrsfs}
\usepackage{amsmath,amsthm}
\def\one{{\hbox{ 1\kern-.8mm l}}}

\def\be{\bar{e}}

\def\be{\begin{equation}}
\def\ee{\end{equation}}

\title{Four-dimensional N=2 Field Theory and Physical Mathematics }

\author{Gregory~W.~Moore\\
\\
 NHETC and Department of Physics and Astronomy,
Rutgers University,\\
Piscataway, NJ 08855--0849, USA\\
\\
{\tt gmoore@physics.rutgers.edu} }

\abstract{We give a    summary of a talk delivered at the
2012 International Congress on Mathematical Physics.
We review d=4, N=2 quantum field theory and some of the exact
statements which can be made about it. We discuss the
wall-crossing phenomenon. An interesting application is
a new construction of hyperk\"ahler metrics on certain
manifolds.  Then we discuss   geometric
constructions which lead to exact results on the BPS spectra
for some d=4, N=2 field theories and on expectation values
of - for example - Wilson line operators. These new constructions
have interesting relations to a number of other areas of
physical mathematics. }

\begin{document}

\section{Introduction}\label{aba:sec1}

The following is a   brief summary of a review talk delivered  at  the
ICMP in Aalborg, Denmark, August 2012. The powerpoint slides are available at \cite{HomePage-1}.
After reviewing some standard material on  d=4, N=2 quantum field theories
we review some work done in a project with Davide Gaiotto and Andy Neitzke
\cite{Gaiotto:2008cd,Gaiotto:2009hg,Gaiotto:2010be,Gaiotto:2011tf,Gaiotto:2012rg,Gaiotto:2012db}.
%
%
A more extensive pedagogical review is in preparation and preliminary
versions are available at \cite{HomePage-2}. Those notes are based on
lectures recently given in Bonn, and the videos are available at
\cite{HomePage-3}. Another brief summary of the construction of hyperk\"ahler
metrics is available at Andy Neitzke's homepage \cite{Andy-HomePage}.

Let us begin with some motivation.
Two important problems in mathematical physics are:

\begin{enumerate}

\item  Given
a quantum field theory (QFT), what is the spectrum of the Hamiltonian,
and how do we compute forces, scattering amplitudes, operator
vev's, etc?

\item  Find solutions to Einstein's equations and find
solutions to the Yang-Mills equations on Einstein manifolds.

\end{enumerate}

The present
work addresses each of these questions within the restricted
context of four-dimensional QFT with N=2 supersymmetry.
Regarding problem 1,   in the past five years there has been much progress in finding
exact results on a portion of the spectrum, the so-called ``BPS spectrum,'' of the
Hamiltonian.  A corollary of this progress is that many
exact results have been obtained for ``line operator'' and ``surface operator''
vacuum expectation values. Regarding problem 2, it turns out that
understanding the BPS spectrum allows one to give very
 explicit constructions of hyperk\"ahler metrics on certain manifolds associated to these
d=4, N=2  field theories. Hyperk\"ahler (HK) manifolds are Ricci flat, and
 hence are solutions to Einstein's equations.
 Moreover,  the results on ``surface operators''  lead
  to a construction of solutions to natural generalizations of the Yang-Mills equations on HK manifolds.
These are hyperholomorphic connections, defined by the condition that the curvature is of type $(1,1)$
in all complex structures. On a 4-dimensional HK manifold a hyperholomorphic connection is the same thing as a self-dual Yang-Mills instanton.

A good development in physical mathematics should open up new questions and directions of research and provide interesting links to other  lines of enquiry. It turns out that solving the above problems leads to interesting relations to
Hitchin systems,  integrable systems,  moduli spaces of flat connections on surfaces, cluster algebras, Teichm\"uller theory and the ``higher Teichm\"uller theory'' of  Fock and Goncharov. The list goes on.
There are many open problems in this field, some of which are mentioned in the conclusions.

\section{d=4 N=2 field theory}

The N=2 super-Poincar\'e algebra is a
$\mathbb{Z}_2$-graded Lie algebra $\mathfrak{S}= \mathfrak{S}^0 \oplus \mathfrak{S}^1$. The even subspace is
 $\mathfrak{S}^0 = iso(1,3) \oplus su(2)_R \oplus \mathbb{R}^2$ where the
second summand on the RHS is a global symmetry known as ``R-symmetry''
and the last summand is central. The odd subspace is in the representation of $\mathfrak{S}^0$
given by   $\mathfrak{S}^1 = [(2,1;2) \oplus (1,2;2)]_{\mathbb{R}}$
where the last subscript is a natural reality condition.  Physicists usually
write the odd generators as $Q_\alpha^A$ and $\bar Q_{\dot \alpha}^A$ where $\alpha, \dot \alpha$
are spin indices and $A=1,2$ is an $SU(2)$ R-symmetry index. The brackets of odd generators are,
using standard Bagger-Wess notation:
\begin{equation}\label{eq:susyalg}
\begin{split}
\{ Q_{\alpha}^{~A}, \bar Q_{\dot \beta B} \} & =    2
\sigma^m_{\alpha\dot\beta}P_m \delta^{A}_{~B} \\
\{ Q_{\alpha}^{~A}, Q_{  \beta}^{~ B} \} & =    2 \epsilon_{\alpha\beta}\epsilon^{AB} \bar Z \\
\{ \bar Q_{\dot\alpha A}, \bar Q_{  \dot \beta  B} \} & =   -2
\epsilon_{\dot\alpha\dot\beta}\epsilon_{AB}   Z .  \\
\end{split}
\end{equation}

Exact N=2 supersymmetry
strongly constrains a QFT.
It constrains the field content, which must be in representations of the supersymmetry algebra,
and it constrains Lagrangians which, for a given field content, typically depend
on far fewer parameters than in the nonsupersymmetric case.
 N=2 also opens up the possibility of ``small'' or ``BPS'' representations
of supersymmetry, over which we have much greater analytical control.

As an example, let us consider $N=2$ supersymmetric Yang-Mills theory (SYM)
for a compact simple Lie group $G$.
In addition to a gauge field $A_\mu^a$, where $\mu=0,1,2,3$ and $a =1, \dots, \dim G$,
 there must be a doublet of gluinos
in the adjoint representation and - very importantly - a pair of real scalar
fields in the \emph{adjoint} representation of the group. These are usually
combined into a complex scalar field with an adjoint index, $\varphi^a$.
In this case the renormalizable Lagrangian is completely determined up to a
choice of Yang-Mills coupling. The Hamiltonian is the sum of the standard
terms and a potential energy term
\begin{equation}\label{eq:CommTerm}
\frac{1}{g^2} \int d^3 x {\rm Tr}([\varphi,\varphi^\dagger])^2.
\end{equation}
This has the important consequence that there is - at least classically -
a \emph{moduli space of vacuum states}. The standard terms of the
Hamiltonian set $E = B =0$
and set $\varphi$ to be a constant in space. The term \eqref{eq:CommTerm}
implies that in the vacuum  $\varphi$ must
be a normal matrix so it can be diagonalized to the form
$\varphi   \in \mathfrak{t}\otimes \mathbb{C}$,
where $\mathfrak{t} $ is a Cartan subalgebra of
the Lie algebra $\mathfrak{g}$ of $G$. Now, a standard set of arguments
(due to Seiberg \cite{Seiberg:1994bp} and Seiberg and Witten \cite{Seiberg:1994rs,Seiberg:1994aj}),
based on the assumption that there is no  anomaly in
supersymmetry and on the strong constraints that $N=2$ supersymmetry
puts on any long-distance effective action, shows that
in fact, \emph{this family of vacua is not lifted in the quantum theory.}
We label the vacua as $\vert \Omega(u)\rangle$, with $u \in {\cal B}:=
\mathfrak{t} \otimes \mathbb{C}/W$, where $W$ is the Weyl group.  For $\mathfrak{g} = su(K)$
the quantum vacua  can be characterized by
the equations
\begin{equation}
\langle \Omega(u) \vert {\rm Tr} \varphi^s \vert \Omega(u) \rangle = u_s \qquad s=2,\dots, K.
\end{equation}
where $u_s$ are complex numbers parametrizing the vacuum.   Informally we can say
\begin{equation}
\langle \Omega(u) \vert \varphi \vert \Omega(u) \rangle  = {\rm Diag} \{ a^1, \dots, a^K \}.
\end{equation}
Physical properties depend on the point $u\in {\cal B}$.

For generic values of $a^1,\dots, a^K$ there is - classically - an unbroken $U(1)^r$
gauge symmetry with $r=K-1$. The low energy theory is therefore described by an N=2
extension of Maxwell's theory, and hence we have
electromagnetic fieldstrengths $F\in \Omega^2(\mathbb{R}^{1,3}; \mathfrak{t})$,
and their superpartners. N=2 supersymmetry constrains the low energy effective
action (LEEA) to be - roughly - of the form
\begin{equation}
S = \int {\rm Im} \tau_{IJ} F^I * F^J + {\rm Re} \tau_{IJ} F^I F^J + {\rm Im}\tau_{IJ} da^I * d \bar a^J + \cdots
\end{equation}
where $\tau_{IJ} = \frac{\theta_{IJ}}{8 \pi } + \frac{4\pi i }{e^2_{IJ}}$ is a
complexified coupling constant. It is a symmetric holomorphic matrix function of the
vacuum parameters $u$.
The theory   contains dyonic particles with both electric and magnetic charges for
the Maxwell fields.
Dirac quantization shows that the electromagnetic charge
 $\gamma$ lies in a symplectic lattice $\Gamma$, with an
 integral antisymmetric form: $\langle \gamma_1, \gamma_2 \rangle \in \mathbb{Z}$.

One of the key features of d=4, N=2 supersymmetry is that one can define
the space of  BPS states.
The Hilbert space of the theory is graded by electromagnetic charge
${\cal H} = \oplus_{\gamma \in \Gamma} {\cal H}_{\gamma}$. Taking the square of
suitable Hermitian combinations of supersymmetry generators and using the algebra
shows that in the sector ${\cal H}_{\gamma}$ there is a Bogomolnyi bound
$E \geq \vert Z_\gamma \vert$ where $Z_{\gamma}$ is the ``central charge'' in the
N=2 supersymmetry algebra \eqref{eq:susyalg}. (On the subspace ${\cal H}_\gamma$
the central charge operator is a $\gamma$-dependent c-number $Z_\gamma$.) The BPS subspace
of the Hilbert space is - by definition - the subspace for which the energy
saturates the Bogomolnyi bound:
\begin{equation}\label{eq:BPS-DEF}
{\cal H}_\gamma^{BPS} := \{ \psi\vert E\psi = \vert Z_\gamma \vert \psi \}.
\end{equation}
The central charge function is linear in $\gamma$, $Z_{\gamma_1+\gamma_2} = Z_{\gamma_1} + Z_{\gamma_2}$,
and is also a holomorphic function of $u$. It turns out that knowing $Z_{\gamma}(u)$ is equivalent
to knowing $\tau_{IJ}(u)$.

So far, everything above follows fairly straightforwardly from general principles.
But how do we actually compute $Z_{\gamma}(u)$ (and hence $\tau_{IJ}(u)$, and hence
the low energy effective dynamics) as a function of $u$?  In a renowned pair of papers
\cite{Seiberg:1994rs,Seiberg:1994aj} Seiberg and Witten showed (for $SU(2)$ N=2 super-QCD)
that $\tau(u)$ can be computed in terms of the periods of a meromorphic differential
form $\lambda$ on a Riemann surface $\Sigma$, both of which depend on $u$. They
therefore showed how to determine the LEEA exactly as a function of $u$. They
also gave cogent arguments for the exact BPS spectrum of the $SU(2)$ theory without quarks.
It was therefore natural to search for the LEEA and the BPS spectrum in other
d=4 N=2 theories. Extensive subsequent work   showed that the Seiberg-Witten
paradigm indeed generalizes to all known solutions for the LEEA of d=4 N=2 theories,
namely, there is a family of Riemann surfaces $\Sigma_u$,
parametrized by the moduli space of vacua, $u \in {\cal B}$,
 together with a meromorphic differential $\lambda_u$ whose periods determine $Z_\gamma(u)$.
The curve $\Sigma_u$ and differential $\lambda_u$ are called the Seiberg-Witten curve
and differential, respectively.
However, to this day, there is no general algorithm for computing the Seiberg-Witten
curve and differential  given an arbitrary d=4, N=2 field theory. It is not even clear, \emph{a priori},
why the Seiberg-Witten paradigm should hold true for such an arbitrary  theory.

One important technical detail in the Seiberg-Witten paradigm should be mentioned here.
There is a complex codimension one
singular locus ${\cal B}^{\rm sing}\subset {\cal B}$ where (BPS) particles become massless.
This invalidates the LEEA, which is only applicable on ${\cal B}^*:= {\cal B} - {\cal B}^{\rm sing}$.
In terms of the Seiberg-Witten curve, some cycle pinches and a period vanishes.
Related to this, the charge lattice has monodromy and hence we should speak of a
local system of charge lattices over ${\cal B}^*$ with fiber at $u$ denoted $\Gamma_u$.

While the LEEA of infinitely many N=2 theories was worked out in the years immediately
following the Seiberg-Witten breakthrough, the BPS spectrum proved to be more
difficult. It was only determined in a handful of cases, using   methods which do
not easily generalize to other theories
\cite{Bilal:1997st,Ferrari:1996sv,Bilal:1996sk,Ferrari:1997gu}.
In the past five years there has been a great deal of progress in understanding
the BPS spectra in an infinite number of N=2 theories. One key element of this
progress has been a much-improved understanding of the ``wall-crossing phenomenon''
to which we turn next.

\section{Wall-crossing 101}

The BPS spaces defined in \eqref{eq:BPS-DEF} are finite dimensional
 representations of $so(3) \oplus su(2)_R$ where $so(3)$ is the spatial rotation
algebra for the little group of a massive particle.
The space \eqref{eq:BPS-DEF} clearly depends on $u$ since $Z_\gamma(u)$ does. However, even the
\emph{dimension} of the space depends on $u$. As in the index theory of Atiyah and Singer,
\eqref{eq:BPS-DEF} is $\mathbb{Z}_2$-graded by $(-1)^{F}$ so there is an index, in
our case a kind of Witten index, which behaves much better as a function of $u$. It is
called the second-helicity supertrace and is defined by
\begin{equation}\label{eq:SecHelST}
\Omega(\gamma) := -\frac{1}{2} {\rm Tr}_{{\cal H}_\gamma^{BPS}} (2J_3)^2 (-1)^{2J_3}
\end{equation}
where $J_3$ is any generator of the rotation algebra $so(3)$. The wall-crossing
phenomenon is the - perhaps surprising - fact that even the \emph{index} can
depend on $u$! Therefore we henceforth write $\Omega(\gamma;u)$.
We hasten to add that the index is piecewise constant in connected open chambers
in ${\cal B}$, separated by real codimension one walls. The essential
physics of this ``wall-crossing''
 is that BPS particles can form boundstates which are themselves BPS.
This phenomenon was first observed in the context of two-dimensional supersymmetric
field theories \cite{Cecotti:1992qh,Cecotti:1992rm}, and it played an important
role in the consistency of the Seiberg-Witten description of pure $SU(2)$ theory
\cite{Seiberg:1994rs}. A quantitative description of
 four-dimensional BPS wall-crossing was first put forward in
\cite{Denef:2007vg}. It is based on a semiclassical picture of BPS
boundstates with BPS constituents.
Indeed, in semiclassical analysis there is a beautiful formula due to Frederik
Denef \cite{Denef:2000nb} which gives the boundstate radius of a boundstate
of two BPS particles of charges $\gamma_1, \gamma_2$ in a vacuum $u$:
\begin{equation}
R_{12}(u) = \langle \gamma_1, \gamma_2 \rangle \frac{\vert Z_{\gamma_1}(u) + Z_{\gamma_2}(u)\vert}{2 {\rm Im} Z_{\gamma_1}(u) Z_{\gamma_2}(u)^*  }.
\end{equation}
The $Z$'s are functions of the moduli $u\in {\cal B}$. We can divide the moduli space of vacua
into regions with $
\langle \gamma_1, \gamma_2 \rangle   {\rm Im} Z_{\gamma_1}(u) Z_{\gamma_2}(u)^*  > 0 $
and
$
\langle \gamma_1, \gamma_2 \rangle   {\rm Im} Z_{\gamma_1}(u) Z_{\gamma_2}(u)^*  < 0 $.
In the latter region the boundstate cannot exist. Now consider a path of vacua $u(t)$
which crosses a ``marginal stability wall,''
\footnote{The reason for the name is that the exact binding energy of the BPS boundstate is
$\vert Z_{\gamma_1+\gamma_2}(u)\vert - \vert Z_{\gamma_1}(u)\vert - \vert Z_{\gamma_2}(u)\vert $,
and hence on the wall, the states are at best marginally bound.}  defined by
\begin{equation}
MS(\gamma_1,\gamma_2) := \{ u \vert Z_{\gamma_1}(u) \parallel Z_{\gamma_2}(u) \qquad \& \qquad
\Omega(\gamma_1;u) \Omega(\gamma_2;u) \not=0 \}.
\end{equation}
 As $u$ approaches this
wall through a region where the boundstate exists the boundstate radius goes to infinity.
We can easily account for the states which leave the Hilbert space. They are:
$\Delta {\cal H} = (J_{12}) \otimes {\cal H}^{BPS}_{\gamma_1} \otimes {\cal H}^{BPS}_{\gamma_2} $
where $(J_{12})$ is the representation of $so(3)$ of dimension $\vert \langle \gamma_1,\gamma_2 \rangle \vert$.
This accounts for the degrees of freedom in the electromagnetic field in the dyonic boundstate. Computing \eqref{eq:SecHelST} for $\Delta{\cal H}$ produces the
``primitive wall-crossing formula'' of \cite{Denef:2007vg}.

However, this is not the full story since when crossing $MS(\gamma_1,\gamma_2)$
other ``multiparticle boundstates'' of total charge $N_1 \gamma_1 + N_2 \gamma_2$
(where $N_1, N_2$ are positive integers) might also decay.
The full wall-crossing formula, which describes all possible bound states which can
form or decay is the ``Kontsevich-Soibelman wall-crossing formula'' (KSWCF)
\cite{KS-1}. Before describing a physical derivation of that formula we first
digress slightly and discuss ``extended operators'' or ``defects'' in quantum field theory,
because   our favorite derivation of the KSWCF uses such ``line defects.''
We should mention, however, that there are other physical derivations
of the KSWCF including \cite{Cecotti:2010qn,Cecotti:2009uf,Manschot:2010qz}.
See also the review \cite{Pioline:2011gf}.

\section{Interlude: Defects in local QFT}

``Extended  operators''   or ``defects''  have been playing an increasingly important role in recent years in quantum field theory. A pseudo-definition would be that    defects are local disturbances  supported on positive codimension submanifolds of spacetime. For example, zero-dimensional defects are just local operators.
Examples of $d=1$ defects are
familiar in gauge theory as Wilson line insertions in the Yang-Mills path integral.
In four-dimensions there are interesting 't Hooft loop defects based on specifying certain singularities in
the gauge field on a linking 2-sphere around a line. Recent progress has relied strongly
on surface defects, where we couple a two-dimensional field theory to an ambient four-dimensional
theory. These 2d4d systems play an important role below.

In general the inclusion of extended objects enriches the notion of QFT.
Even in the case of topological field theory, the usual formulation of Atiyah and Segal is
enhanced to ``extended TQFTs'' leading to beautiful relations with higher category theory.
We will not need that mathematics here, but the interested reader might consult
\cite{Kapustin:2010ta,Lurie:Expository} for further information.

\section{Wall Crossing 102}

We will now use   line defects to produce a  physical derivation of the KSWCF.
This is an argument which appears in more detail in
\cite{Gaiotto:2010be,Andriyash:2010yf,Andriyash:2010qv}.  We consider line defects
sitting at the origin of space, stretching along the (Euclidean or Lorentzian) time
direction and preserving a linear combination of supersymmetries of the form
$Q + \zeta \bar Q$ where $\zeta$ is a phase. We generally denote such line defects by $L_\zeta$.
 A good example is the supersymmetric
extension of the Wilson line in N=2 SYM:
\begin{equation}\label{eq:WilsonLine}
L_\zeta = \exp \int_{\mathbb{R}_t \times \{ \vec 0 \} } \left( \frac{\varphi}{2\zeta} + A + \frac{\zeta}{2} \bar\varphi\right).
\end{equation}
For any line defect $L_\zeta$ the Hilbert space, as a representation of the
 superalgebra, is modified to ${\cal H}_{L_\zeta}$ and in the N=2 theories
it is still graded by $\Gamma$, or rather by a $\Gamma$-torsor:
\begin{equation}
{\cal H}_{\gamma} = \oplus_{\gamma \in \Gamma + \gamma_0} {\cal H}_{L_\zeta,\gamma}.
\end{equation}
The physical picture of the charge sector $\gamma$ is that we have effectively inserted an
infinitely heavy BPS particle of charge $\gamma$ at the origin of space. The \emph{framed BPS states} are
states in ${\cal H}_{L_\zeta,\gamma}$ which saturate a modified BPS bound.
This bound applies to
these modified Hilbert spaces and is  $E\geq - {\rm Re}(Z_\gamma/\zeta)$. Once again we can define
a \emph{framed BPS index}:
\begin{equation}
{\overline{\underline{\Omega}}}(L_\zeta;\gamma) := {\rm Tr}_{{\cal H}^{BPS}_{L_\zeta,\gamma} }(-1)^{2J_3}.
\end{equation}
If we consider line defects of type $L_\zeta$ then these framed BPS indices will be
piecewise constant in $\zeta$ and $u$ but again exhibit wall-crossing, this time
across ``BPS walls'' defined   by
\begin{equation}
W_\gamma := \{ (u,\zeta) \vert Z_{\gamma}(u)/\zeta < 0\qquad \& \qquad \Omega(\gamma;u)\not=0 \}.
\end{equation}
The physical significance of these walls
is that when $(u,\zeta)$ are close to the wall there is a
subsector of ${\cal H}^{BPS}_{L_\zeta}$ which is described -semiclassically -
by states in which a collection of BPS particles with charges of the form   $n\gamma$,
with $n>0$,  is bound to a defect in
charge sector $\gamma_c$ to make a framed BPS state with boundstate radius
\begin{equation}
r_\gamma = \frac{\langle \gamma, \gamma_c \rangle}{2 {\rm Im} Z_{\gamma}(u)/\zeta }.
\end{equation}
In fact since BPS particles of charge $n \gamma$ for $n>0$ can bind in arbitrary
numbers to the core defect, (this is possible since they feel no relative force)
 there is an
entire Fock space of boundstates of these so-called ``halo particles.''
When crossing the wall this entire Fock space appears or disappears in the
framed Hilbert space. Exactly the same physical picture underlies the
``semi-primitive wall-crossing formula'' of  \cite{Denef:2007vg}.

An elegant way to express this wall-crossing mathematically is the following.
Introduce the framed BPS degeneracy generating function
\begin{equation}
F(L):= \sum_{\gamma} {\overline{\underline{\Omega}}}(L;\gamma)  X_\gamma
\end{equation}
where $X_{\gamma_1} X_{\gamma_2} = (-1)^{\langle \gamma_1, \gamma_2 \rangle} X_{\gamma_1+\gamma_2}$
generate the twisted algebra of functions on an algebraic torus $\Gamma^*\otimes \mathbb{C}^*$.
When crossing a BPS
wall $W_\gamma$ the charge sectors of the form $\gamma_c + N \gamma$ gain or lose a
Fock space factor:
\begin{equation}
X_{\gamma_c} \to (1 - (-1)^{\langle \gamma_1, \gamma_2 \rangle} X_{\gamma})^{\langle \gamma , \gamma_c \rangle\Omega(\gamma) }X_{\gamma_c}
\end{equation}
Once again the factor $\langle \gamma, \gamma_c\rangle$ accounts for degrees of freedom in the
electromagnetic field. Since the wall-crossing factor depends on $\gamma_c$,
the change of $F(L)$ across a BPS wall $W_\gamma$ is given by the action
of a \emph{differential operator}: $F(L) \to K_\gamma^{\Omega(\gamma)} F(L)$ where
\begin{equation}
K_\gamma = (1-(-1)^{D_{\gamma}} X_\gamma)^{D_\gamma}
\end{equation}
 and $D_\gamma X_\rho = \langle \gamma, \rho \rangle X_\rho$.
We now consider a point $u_*$ on the marginal stability wall $MS(\gamma_1,\gamma_2)$.
The intersection of the BPS walls $W_{r_1 \gamma_1 + r_2 \gamma_2}$ which go
through $u_*$ and have   $r_1 r_2 \geq 0$ defines a complex codimension one locus
in ${\cal B}^*$.  Now consider two small paths linking this locus with one path in a region
where ${\rm Im}Z_1 \bar Z_2>0$ and the other in the region ${\rm Im}Z_1 \bar Z_2<0$.
On the one hand the generating function $F(L)$ is
is well-defined at the endpoints of the paths: There is no
monodromy in the continuous evolution  of $F(L)$ around the loop. On the other hand transport
of $F(L)$ along the path leads to a sequence of transformations by $K_\gamma^{\Omega(\gamma)}$ each time the point $u(\tau)$ on the path goes through a wall $W_\gamma$.
These two statements together
\footnote{plus an important detail that there be ``sufficiently many'' line defects}
imply  the KSWCF:
\begin{equation}\label{eq:KSWCF}
\prod_{\nearrow} K^{\Omega(r_1 \gamma_1 + r_2 \gamma_2; - )}_{r_1 \gamma_1 + r_2 \gamma_2} =
\prod_{\searrow} K^{\Omega(r_1 \gamma_1 + r_2 \gamma_2; + )}_{r_1 \gamma_1 + r_2 \gamma_2}
\end{equation}
where the product is over pairs of nonnegative integers $(r_1,r_2)$.
The product with   $\nearrow$ is ordered with  $r_1/r_2$ increasing from left to right,
while that with $\searrow$ is
ordered with  $r_1/r_2$ decreasing from left to right. The $\pm $ in $\Omega$ refers to the
 BPS degeneracies on either side of the wall.
 Knowing the $\Omega(r_1 \gamma_1 + r_2 \gamma_2; - )$
we compute the LHS of the equation. Given an ordering of the $K_\gamma$ factors
there is a unique factorization of this product of the form in \eqref{eq:KSWCF}.
Hence, given $\Omega(r_1 \gamma_1 + r_2 \gamma_2; - )$ the
$\Omega(r_1 \gamma_1 + r_2 \gamma_2; + )$ are uniquely determined. Equation
\eqref{eq:KSWCF}   is therefore a wall-crossing formula.

Two examples serve to illustrate the theory well. If $\Gamma = \gamma_1 \mathbb{Z} \oplus
\gamma_2 \mathbb{Z}$ and $\langle \gamma_1, \gamma_2 \rangle = +1$ then
\begin{equation}
K_{\gamma_1} K_{\gamma_1} = K_{\gamma_1} K_{\gamma_1 + \gamma_2} K_{\gamma_2}.
\end{equation}
This identity is easily verified. It is related to consistency of simple
superconformal field theories (``Argyres-Douglas theories'') as well as to
coherence theorems in category theory, 5-term dilogarithm identities, and
a number of other things. Our second example again takes $\Gamma = \gamma_1 \mathbb{Z} \oplus
\gamma_2 \mathbb{Z}$ but now with  $\langle \gamma_1, \gamma_2 \rangle = +2$. Then
\begin{equation}
K_{\gamma_2} K_{\gamma_1} = \Pi_L K_{\gamma_1+\gamma_2}^{-2} \Pi_R
\end{equation}
\begin{equation}
\begin{split}
\Pi_L & =\prod_{n=0\nearrow \infty} K_{(n+1)\gamma_1 + n \gamma_2} =
 K_{\gamma_1}  K_{2\gamma_1 + \gamma_2}\cdots \\
\Pi_R & =\prod_{n=\infty \searrow 0 } K_{n \gamma_1 + (n+1) \gamma_2} =
\cdots K_{\gamma_1 + 2 \gamma_2} K_{\gamma_2} \\
\end{split}
\end{equation}
This identity perfectly captures the wall-crossing of the BPS spectrum
found in the original example of Seiberg and Witten \cite{Seiberg:1994rs},
a remark due to Frederik Denef. The corresponding
identities for the cases $\langle \gamma_1, \gamma_2 \rangle \not= 0,1,2$
are considerably wilder.

We stress that this is only half the battle. The wall-crossing formula only
describes the \emph{change} of the BPS spectrum across a wall of marginal stability.
It does \emph{not} determine the BPS spectrum! For a certain (infinite) class
of N=2 theories - the theories of class S - we can do better and give an algorithm
to determine  the BPS spectrum, as we describe below.

\section{Reduction to three dimensions and hyperk\"ahler geometry}

Interesting relations to hyperk\"ahler geometry emerge when we
compactify N=2 theories on a circle of radius $R$. At energy
scales much lower than $1/R$ the theory is described by
a supersymmetric sigma model with target space $\cal M$
which comes with a natural torus fibration over ${\cal B}$
\cite{Seiberg:1996nz}.
The presence of 8 supersymmetries means that $\cal M$ must
carry a hyperk\"ahler metric.  In the large $R$ limit
this metric can be easily solved for, but at finite values
of $R$ there are nontrivial quantum corrections.
The idea of the construction of \cite{Gaiotto:2008cd} is
to find a suitable set of functions on the twistor space
of $\cal M$ from which one can construct the metric.
The required functions turn out to be solutions
to an explicit integral equation closely resembling Zamolodchikov's
thermodynamic Bethe ansatz.

The low energy three-dimensional sigma model has scalar fields $a^I(x) \in \cal B$
descending from the scalars in four dimensions, as well as
two periodic scalars $\theta_e^I(x)$ and $\theta_{m,I}(x)$ for each
dimension $I=1,\dots, r$ of $\mathfrak{t}$. We can think of
$\theta_e^I(x) = \oint_{S^1} A $ as the ``Wilson loop scalar''
and $\theta_{m,I}(x)$ as an electromagnetic dual scalar, coming
from dualization of the three-dimensional gauge field. This leads to
a picture of the target space as a fibration by tori, whose
generic fiber is $\Gamma^*_u \otimes \mathbb{R}/2\pi \mathbb{Z}$.
In this way we find a direct relation to integrable systems.  The semiflat metric on
this space is computed in a straightforward way from the reduction
of the four-dimensional LEEA of Seiberg-Witten and leads to
\begin{equation}
g^{\rm sf} = da^I R {\rm Im}\tau_{IJ} d \bar a^J + \frac{1}{R} dz_I ({\rm Im} \tau)^{-1,IJ} d \bar z_J
\end{equation}
where $dz_I = d \theta_{m,I} - \tau_{IJ} d \theta_e^J $.  This metric will receive
quantum corrections.

The best way to approach the quantum corrections is to form the twistor space
$Z:= {\cal M} \times \mathbb{C} P^1$ which comes with a fibration $p:Z \to \mathbb{C} P^1$.
A theorem of Hitchin says that putting a hyperk\"ahler metric on $\cal M$ is
equivalent to putting holomorphic data on $Z$ so that the fiber  $p^{-1}(\zeta)$
 above a point $\zeta \in \mathbb{C} P^1$  is $\cal{M}$ in complex structure $\zeta$.
 Moreover there is a holomorphic 2-form form $\varpi \in \Omega^2_{Z/\mathbb{C} P^1} \otimes {\cal O}(2)$
 which restricts on each fiber to the holomorphic symplectic form $\varpi_\zeta$
 of ${\cal{M}}^\zeta$ and which, as a function of $\zeta$, has
  a three-term Laurent expansion in $\zeta \in \mathbb{C}^*$:
 \begin{equation}
 \varpi_\zeta = \zeta^{-1} \omega_+ + \omega_3 + \zeta \omega_-.
 \end{equation}
 Here $\omega_+$ is a holomorphic $(2,0)$ form in complex structure $\zeta=0$ and $\omega_3$ is
 the K\"ahler form of the metric.

The strategy of the construction is to find $\varpi_\zeta$ by covering $\cal M$ with coordinate
charts of the form
\begin{equation}
{\cal U} = \Gamma^* \otimes \mathbb{C}^* \cong
\underbrace{\mathbb{C}^* \times \cdots \mathbb{C}^*}_{2r}.
\end{equation}
The algebraic torus has a canonical set of ``Darboux functions'' $Y_\gamma$ given (up to a
sign)
\footnote{The sign is determined by a mod-two quadratic refinement of the
intersection form.}  by evaluation
with $\gamma \in \Gamma$ and satisfying $Y_{\gamma_1} Y_{\gamma_2} = (-1)^{\langle \gamma_1, \gamma_2 \rangle} Y_{\gamma_1 + \gamma_2}$. In terms of these we can write a canonical holomorphic symplectic
form $\varpi_T$ by choosing a basis $\{ \gamma_i \}$ for $\Gamma$ and writing
$\varpi_T = C^{ij} d \log Y_{\gamma_i} \wedge d \log Y_{\gamma_j} $ where $C_{ij}$ is the
symplectic form of $\Gamma$ in that basis.  Thus, we seek suitable holomorphic maps
\begin{equation}
{\cal Y}: {\cal U} \times \mathbb{C}^* \rightarrow \Gamma^* \otimes \mathbb{C}^*,
\end{equation}
where the second factor in the domain is the twistor sphere stereographically projected,
such that $\varpi_\zeta = \cal Y^*(\varpi_T)$ has a 3-term Laurent expansion.

For the semiflat metric one can solve for these ``Darboux functions'' in a straightforward
way to obtain
\begin{equation}
{\cal Y}_\gamma^{\rm sf} = \exp \left[ \pi R \zeta^{-1} Z_{\gamma} + i \theta_\gamma + \pi R \zeta \bar Z_\gamma\right]
\end{equation}
where $\theta_\gamma$ is a   linear combination of $\theta_e^I$ and $\theta_{m,I}$ such that
${\cal Y}_{\gamma}^{\rm sf}{\cal Y}_{\gamma'}^{\rm sf}= (-1)^{\langle \gamma, \gamma' \rangle}
{\cal Y}_{\gamma+\gamma'}^{\rm sf}$.
The goal, then,  is to find the quantum corrections: $\cal Y_\gamma = \cal Y_{\gamma}^{\rm sf} \cal Y_{\gamma}^{\rm quant. corr.}$.  The desired properties of the exact functions ${\cal Y}_\gamma(u,\theta_e,\theta_m;\zeta)= {\cal Y}^*(Y_\gamma)$ leads to a list of conditions which are equivalent to a Riemann-Hilbert problem
in the complex $\zeta$-plane. This RH problem is then solved by the integral equation
\begin{equation}\label{eq:TBA}
\log {\cal Y}_\gamma = \log {\cal Y}_{\gamma}^{\rm sf} + \frac{1}{4\pi i } \sum_{\gamma_1\in \Gamma} \Omega(\gamma_1;u)
\langle \gamma_1, \gamma \rangle \int_{\ell_{\gamma_1}} \frac{d\zeta_1}{\zeta_1} \frac{\zeta_1 + \zeta}{\zeta_1 - \zeta}
\log(1- {\cal Y}_{\gamma_1}(\zeta_1) ),
\end{equation}
where $\ell_{\gamma}$ is the projection, at fixed $u$, of $W_\gamma$ to $\mathbb{C}^*$.
This equation can be solved by iteration for sufficiently large $R$ and for sufficiently
tame BPS spectrum. (We expect a typical field theory to be ``tame,'' but a typical
black hole spectrum will definitely not be tame. New ideas are needed to apply these
techniques to supergravity.  See \cite{Alexandrov:2011ac} for the state of the art.)
The ${\cal Y}_\gamma(\zeta)$ jump discontinuously across the BPS walls in the $\zeta$ plane,
but that discontinuity is a symplectic transformation, so that $\varpi_\zeta$ is continuous.
Note well that the   BPS spectrum is an important input into \eqref{eq:TBA}. As we have
explained, it is discontinuous in $u$ because of wall-crossing. Nevertheless,
across walls of marginal stability in ${\cal B}$ the metric is continuous, thanks to the
KSWCF. Indeed, one can reverse the logic: Physically no discontinuity in the metric
is expected across marginal stability walls, and therefore we \emph{derive}
the KSWCF \cite{Gaiotto:2008cd}.

The ``Darboux functions'' ${\cal Y}_\gamma$ have other useful applications. For example,
they can be used to write exact results for expectation values for line defects.
For example, wrapping a line defect of type $\zeta$ around the compactification circle
produces a \emph{local operator} ${\rm tr} L_\zeta$ in the three-dimensional sigma model.
The vacua of the model are points $m \in {\cal M}$. In \cite{Gaiotto:2010be} it is
argued that the vev of this operator in the vacuum $m$ is
\begin{equation}\label{eq:DarbouxExpansion}
\langle {\rm tr}  L_{\zeta} \rangle_{m} = \sum_{\gamma}
{\overline{\underline{\Omega}}}(L_{\zeta} ;\gamma){\cal Y}_{\gamma}(m,\zeta).
\end{equation}
A related formula leads to a natural deformation quantization of the
algebra of holomorphic functions on ${\cal M}^\zeta$. An extension of the
above integral equations leads to a construction of hyperholomorphic
connections on ${\cal M}$ \cite{Gaiotto:2011tf}.

\section{Theories of class S}

We now turn to a rich set of examples of d=4, N=2 theories, known as
the ``theories of class S.'' The ``S'' is for ``six'' because these are
N=2 theories which descend from six-dimensional theories. In these theories
many physical quantities have elegant descriptions in terms of Riemann surfaces
and flat connections.

The  construction is based on an important claim arising from string theory,
namely, that there is a family of stable interacting UV-complete field theories with
six-dimensional (2,0) superconformal symmetry \cite{Witten:1995zh,Strominger:1995ac,Seiberg:1997ax}.
These theories have not yet been constructed - even by physical standards - but some characteristic
properties of these hypothetical theories can be deduced from their relation to
string theory and M-theory. For a review based on this philosophy
see \cite{Gaiotto:2010be}, \S 7.1 or the preliminary notes \cite{HomePage-2} .

In order to construct theories of class S we begin with such a nonabelian (2,0)
theory in six dimensions, $S[\mathfrak{g}]$, where $\mathfrak{g}$ is a simple and
simply laced compact real Lie algebra. The theory has half-BPS codimension two
defects $D$. We compactify the theory on a Riemann surface $C$, referred to as
 the ``ultraviolet curve.'' The surface $C$ has marked points
  $\mathfrak{s}_n$ and we put defects $D_n$ at
$\mathfrak{s}_n$. Then we partially topologically twist, by embedding $so(2)$
into the $so(5)_R$ R-symmetry of the $(2,0)$ superconformal algebra and identifying
with the algebra of the structure group of the tangent bundle $TC$.
The resulting theory - at least formally - only depends on the conformal class of the
metric through the overall area. In the limit
where the area of $C$ shrinks to zero\footnote{There can be subtleties in taking this
limit if there are too few or nongeneric defects \cite{Gaiotto:2011xs}.} we obtain a four-dimensional
quantum field theory denoted $S(\mathfrak{g},C,D)$. This construction goes back to
\cite{Witten:1997sc}. It has a dual version given by geometric engineering in
\cite{Klemm:1996bj}. The topological twisting, defects, and relations to
Hitchin systems were given in \cite{Gaiotto:2009hg}. The construction
 was then further developed in a brilliant paper of Gaiotto \cite{Gaiotto:2009we}.

Although it will not play any direct role in the rest of our story,
we must digress briefly to comment on one important insight from \cite{Gaiotto:2009we}
which we regard as very deep. Defects have global symmetries.
The theory $S(\mathfrak{g},C,D)$ has a global symmetry group which
includes a product over $n$ of the global symmetries of $D_n$.
For suitable defects  $D_n$ (known as ``full defects'')  the global symmetry is just a
compact group  $G$   with Lie algebra $\mathfrak{g}$.
Therefore,
if we have two Riemann surfaces $C_L$ and $C_R$ with collections of
defects $D_L$ and $D_R$ containing at least one such full defect in
 each collection we can consider the global symmetry factor $G$ from
each surface and gauge it with parameter $\tau$. This produces a
new four-dimensional theory $S(\mathfrak{g},C_L, D_L) \times_{G,\tau}
S(\mathfrak{g},C_R, D_R)$. On the other hand, given marked points $\mathfrak{s}_L$
and $\mathfrak{s}_R$ on $C_L$ and $C_R$, respectively we can choose
local coordinates $z_L$ and $z_R$ and form a new glued Riemann surface
$C_L\times_q C_R$ by identifying $z_L z_R = q$. Therefore we can form
 a new quantum field theory $S(\mathfrak{g}, C_L\times_q C_R, D_{LR})$,
where $D_{LR}$ is the union of the sets of left and right defects, omitting
the two associated with the glued marked points. Gaiotto's conjecture is that
 these two four-dimensional N=2 theories are in fact the same, provided
 we identify $q=e^{2\pi i \tau}$.
 Many beautiful results flow from this observation. It is probably the
 fundamental reason for the AGT conjecture \cite{Alday:2009aq},
 although that intuition has not yet been made very precise. One precise
 mathematical version of this phenomenon,   related to Higgs branches
 of vacua of these theories,   is described in \cite{Moore:2011ee}.

Most ``natural'' d=4, N=2 theories are of class S. For example, the N=2
extension of $SU(K)$ Yang-Mills coupled to quark flavors in the
fundamental representation is of class S.
Moreover, there are infinitely many theories of class S with no known
Lagrangian description such as the Argyres-Douglas theories described in
\cite{Gaiotto:2009hg} or the higher rank superconformal fixed points
associated with three-punctured spheres (``trinion theories'')
which were discovered in \cite{Gaiotto:2009we}.

One of the nicest properties of these theories is their close relation to
Hitchin systems. This can be seen very directly \cite{Gaiotto:2009hg} by considering the compactification
of the $(2,0)$ theory on $S^1 \times C$. Compactifying in either order, and
using the crucial fact that the long distance dynamics of the $(2,0)$
theory on a circle of radius $R$ is described by nonabelian five-dimensional
SYM with $g_{YM}^2 \sim R$, shows that
for these theories ${\cal M}$ can be identified with the moduli space
of solutions to Hitchin's equations for a gauge connection and ``Higgs field''
on $C$:
\begin{align}
F + R^2 [\varphi , \bar\varphi] & = 0, \label{eq:hitchin-1} \\
 \bar\partial_A \varphi := d\bar z\left( \partial_{\bar z} \varphi  + [ A_{\bar z}  ,\varphi]\right) & = 0, \label{eq:hitchin-2} \\
 \partial_A \bar\varphi := dz\left( \partial_z \bar\varphi  + [ A_z,\bar\varphi] \right)& = 0. \label{eq:hitchin-3}
\end{align}
Here $A$ is a unitary connection on an Hermitian vector bundle over $C$,
$\varphi$ is an  adjoint valued  $(1,0)$-form  field and $\bar\varphi = \varphi^{\dagger}$
is its Hermitian conjugate.  A defect $D_n$ at $\mathfrak{s}_n$ induces a singularity  in the Higgs field
of the form
\begin{equation}
\varphi \sim \frac{\mathfrak{r}_n}{z^{\ell_n}} dz  + \cdots  \qquad \ell_n \geq 1
\end{equation}
where $z$ is a local coordinate near $\mathfrak{s}_n$ and $\ell_n$ and $\mathfrak{r}_n$
depend on $D_n$.  The physics depends
on $\ell_n$ and $\mathfrak{r}_n $ in a way which is still being understood.
The state of the art is summarized nicely in \cite{Chacaltana:2012zy}.
Later we will use an important connection to complex flat connections:
If $(\varphi, A)$ solve the Hitchin
equations   and $\zeta \in \mathbb{C}^*$ then
\begin{equation}\label{eq:Azeta}
{\cal A}(\zeta) := \frac{R}{\zeta}\varphi + A + R \zeta \bar \varphi
\end{equation}
is flat: $d{\cal A} + {\cal A} \wedge {\cal A} =0$. Conversely given
a family of such complex flat connections, ${\cal A}(\zeta)$ with a
three-term Laurent expansion,  $(\varphi, A)$ solve the Hitchin equations.

We now state how the Seiberg-Witten curve and differential, the charge lattice,
the Coulomb branch, the BPS states, and a natural class of line and surface defects
can all be formulated geometrically in terms of the geometry and topology of
the UV curve $C$ and its associated flat connection ${\cal A}$.

First, the Seiberg-Witten curve is simply
\begin{equation}
\Sigma: = \{ \lambda\vert \det(\lambda - \varphi) = 0 \} \subset T^*C
\end{equation}
and it inherits a canonical differential $\lambda$ which serves as
the Seiberg-Witten differential. For $\mathfrak{g}=su(K)$, the map
$\pi: \Sigma \to C$ is a $K$-fold branched cover and
this equation can be written as
\begin{equation}
\lambda^K + \lambda^{K-2}\phi_2 + \cdots + \phi_K =0
\end{equation}
where $\phi_j$ are meromorphic $j$-differentials with prescribed singularities
at $\mathfrak{s}_n$. From this we deduce that ${\cal B}:= \{ u= (\phi_2,\dots, \phi_K) \}$
is a torsor for a space of meromorphic differentials on $C$. Similarly the
local system of charges is\footnote{Actually, $\Gamma$ is a subquotient. We will
 ignore this subtlety in this brief review for simplicity.}
 $\Gamma = H_1(\Sigma;\mathbb{Z})$.

The geometric formulation of BPS states in these theories goes back
to \cite{Klemm:1996bj,Mikhailov:1997jv,Mikhailov:1998bx}. We take
$\mathfrak{g}=su(K)$. We label
the sheets of the covering $\pi: \Sigma \to C$ by $i,j=1,\dots, K$.
We define a \emph{WKB path of phase $\vartheta$} to be a local solution
of a differential equation on $C$:
\begin{equation}
\langle \lambda_i - \lambda_j , \partial_t \rangle= e^{i \vartheta}
\end{equation}
where $i,j$ is an ordered pair of sheets of the covering.
\footnote{For $su(2)$ a WKB path is just the trajectory of
a quadratic differential $\phi_2$. These have been  widely
studied in the mathematical literature. We think the generalization
to $K>2$ is very rich and interesting. } The marked points $\mathfrak{s}_n$
act like attractors for the WKB paths. Therefore,
 for generic initial point and generic $\vartheta$
both ends of a WKB path tend to such marked points.   One
interesting exception is a WKB  path beginning on a branchpoint.
But once again,  for a generic $\vartheta$, the other end of such a WKB path
terminates on a marked point. The adjective \emph{generic} used above
is quite important. For  special values of $\vartheta$ we can have \emph{string
webs}. These are closed WKB paths, or connected graphs with all
endpoints (if any) on branch points. The graphs comprising string webs are allowed to
have trivalent vertices, known as \emph{string junctions}. The three legs of the
string junction   consist of ingoing  $ij$ and $jk$ WKB paths with an outgoing
 $ik$ WKB path.

There is a geometrical construction, beginning with the six-dimensional $(2,0)$ theory
and  any closed continuous path $\wp\subset C$, which produces a  line defect
in $S(\mathfrak{g}, C, D)$.   The construction also
depends on an angle $\vartheta$ so we denote these line defects as $L_{\wp,\vartheta}$.
The ``Darboux expansion'' \eqref{eq:DarbouxExpansion} together with a
relation of ${\cal Y}_\gamma$ to Fock-Goncharov coordinates on moduli spaces
of flat connections allows us to write physically interesting exact results
for expectation values of such line defects. For example, for   N=2
SU(2) SYM the vev of the Wilson line operator  \eqref{eq:WilsonLine} wrapped around a Euclidean
time circle of radius $R$ is, exactly,
\begin{equation}
\langle {\rm tr} L_{\zeta} \rangle = \sqrt{{\cal Y}_{\gamma_e}} +
\frac{1}{ \sqrt{{\cal Y}_{\gamma_e}} } + \sqrt{{\cal Y}_{\gamma_e + \gamma_m } }.
\end{equation}
The first two terms, with ${\cal Y} \to {\cal Y}^{\rm sf}$ give the naive
semiclassical approximation. The third term  is exponentially small.
This, together with the   the full sum of
instanton corrections to ${\cal Y}^{\rm sf}$ give the complete set of the quantum corrections.
It is not an accident that this expression bears a very strong relation to the expectation value
of a length operator in quantum Teichm\"uller theory \cite{Teschner:2005bz}.

There is one last construction for theories of class S we will need
\cite{Alday:2009fs,Gaiotto:2009fs,Gaiotto:2011tf}. This
is the canonical surface defect $\mathbb{S}_z$ associated with any point $z\in C$.
It is a 1+1 dimensional QFT located at, say, $x^1=x^2=0$ in four-dimensions
and coupled to the ambient four-dimensional theory $S(\mathfrak{g},C,D)$.
The main fact we need about this theory is that (so long as $z$ is
not a branch point of $\pi: \Sigma \to C$)  it has massive vacua in
1-1 correspondence with the preimages $z^{(i)}\in \Sigma$ of $z$ under $\pi$.
Moreover, in the theory $\mathbb{S}_z$ there are solitons interpolating
between vacua $z^{(i)}$ and $z^{(j)}$ for $i\not=j$. These two-dimensional
solitons are represented
geometrically by \emph{open string webs} which are defined as above for string webs
but one end of the graph must end at $z$.

\section{Spectral Networks}

As we have emphasized,  the KSWCF by itself does not give us the BPS spectrum.
For theories of class S we can solve this problem, at least in principle,
with the technique of spectral networks \cite{Gaiotto:2012rg}.
 Spectral networks are combinatorial
objects associated to a branched covering of Riemann surfaces $\pi: \Sigma \to C$.
They are networks ${\cal W}_{\vartheta} \subset C$
defined by the physics of two-dimensional solitons on the surface defect $\mathbb{S}_z$. Segments  in the network are constructed from WKB paths of phase $\vartheta$  according to   local rules given in \cite{Gaiotto:2012rg}.
There can be interesting discontinuous changes in ${\cal W}_\vartheta$ as $\vartheta$ is varied.
Some amusing movies of these morphisms of spectral networks can be
viewed at A. Neitzke's
homepage \cite{Movies}. The essential jumps of the spectral networks happen precisely at those values
of $\vartheta$  which are   the phases of central charges of four-dimensional
BPS states. Indeed, one can write very explicit formulae for the BPS degeneracies
 $\Omega(\gamma;u)$ in the theories $S(su(K), C,D)$ in terms of the combinatorics of
 the change of the spectral network ${\cal W}_{\vartheta}$ as $\vartheta$
 passes through such a critical value  \cite{Gaiotto:2012rg}.
Spectral networks have at least three nice applications to mathematics.

 The first application comes from specializing the construction of the
 hyperholomorphic connections mentioned above to the theories of class S.
  The extra integral equations
 in this case are generalizations of the Gelfand-Levitan-Marchenko
 equation of integrable systems theory and give in principle a way to
 construct explicit solutions to Hitchin's equations on $C$ \cite{Gaiotto:2011tf}.

 A second, closely related,  application is that
 they provide the essential data needed to construct a holomorphic symplectic
 ``nonabelianization map''
 \begin{equation}
 \Psi_{\cal W}: {\cal M}(\Sigma, GL(1);\mathfrak{m}) \to {\cal M}_F(C, GL(K),\mathfrak{m})
 \end{equation}
 which maps flat $GL(1,\mathbb{C})$ connections on $\Sigma$ with specified monodromy
 $\mathfrak{m}_n^{(i)}$ around the lifts $\mathfrak{s}_n^{(i)}$ to flat $GL(K,\mathbb{C})$
 connections on $C$ with specified conjugacy classes of monodromy and flag structure
 at $\mathfrak{s}_n$. The map depends on a choice ${\cal W}$ of
 spectral network. The holonomies of the flat connection $\nabla^{\rm ab}$
 such that $\Psi_{\cal W}(\nabla^{\rm ab}) = \nabla$ define a set of
 holomorphic functions
 ${\cal Y}_\gamma =  \exp\oint_{\gamma}  \nabla^{\rm ab}$
 in a chart ${\cal U}_{\cal W}\subset{\cal M}_F(C, GL(K),\mathfrak{m})$
 where $\Psi_{\cal W}$ is invertible. Choosing a basis for $\Gamma$ we then
 obtain a local coordinate system in the chart  ${\cal U}_{\cal W}$.
 These coordinates  depend on
 the spectral network. Comparing the coordinates across two charts, where
 ${\cal W}$ and ${\cal W}'$ are related by a simple morphism associated with a
 four-dimensional BPS state, leads to a change of coordinates closely resembling
 a cluster transformation. The coordinates ${\cal Y}_{\gamma}$
 thereby provide a system of coordinates on moduli spaces
 of flat connections which appear to generalize the cluster coordinates of
 Thurston, Penner, Fock, and Fock and Goncharov. For the case $K=2$, and
 in some nontrivial examples with $K>2$,  they coincide with coordinates
 defined by Fock and Goncharov, as shown in \cite{Gaiotto:2009hg,Gaiotto:2012db},
 respectively.

 The third application is to WKB
 theory. The $K\times K$ matrix equation on $C$:
 \begin{equation}
 \left( \frac{d}{dz} + {\cal A} \right) \psi = 0
 \end{equation}
  is an ODE generalizing the Schrodinger equation
  (which occurs with $K=2$). If ${\cal A}$ is of the
  form \eqref{eq:Azeta} then we can study the $\zeta \to 0$
  (or $\zeta \to \infty$) asymptotics at fixed $(\varphi,A)$.  The extension from $K=2$
  to $K>2$ is nontrivial. The spectral networks can be interpreted
  as the Stokes lines for this problem \cite{Gaiotto:2012rg}.

  \section{Conclusions}

  In conclusion, we have a good physical understanding of wall-crossing,
  and some improved understanding of how to compute the BPS spectrum,
  at least for theories of class S. Compactification on a circle leads to a new
  construction of hyperk\"ahler metrics and hyperholomorphic connections.
  As a by-product we find many new and nontrivial results on line and surface
  defects and their associated BPS spectra, again in theories of class S.

  Among the many open problems and future directions in this field we mention
  but a few.
  One problem is to make the spectral network technique more effective.
  Another is to give a direct relation to other recent works which have made
  important progress in the computation of the BPS spectra of N=2 theories, e.g. 
  through BPS quivers 
  \cite{Denef:2002ru,Alim:2011ae,Alim:2011kw}, or 
  geometric engineering  \cite{Cecotti:2012jx,GeomEng-FramedBPS}. One natural question is whether it is
  possible to classify d=4, N=2 theories, and whether the theories of class S
  constitute - in some sense - ``most''  N=2 theories. Another interesting
  problem is whether the construction of hyperk\"ahler metrics described
  above can be used to produce explicit metrics on - say - K3 surfaces.
  In another direction, the independence of the twisted theory from the
  K\"ahler class of the metric on $C$, together with the Gaiotto gluing conjecture mentioned
  above implies that, in some sense,  $(2,0)$ theories can be used to
  define a notion of ``two-dimensional conformal field theories valued in
  four-dimensional theories.'' It would be interesting to make that sense
  mathematically precise.

  Finally, there are three broader points we would like to stress.
  First: Seiberg and Witten's  breakthrough in 1994 opened up
   many interesting problems.  Some were quickly solved, but some, related
   to the computation of  the BPS spectrum,  remained stubbornly open.
The past five years has witnessed a renaissance of the subject, with a much deeper understanding of the BPS spectrum and of the line and surface defects in these theories. Second:
This progress has involved nontrivial and surprising connections to other aspects of physical mathematics
including hyperk\"ahler geometry, cluster algebras, moduli spaces of flat connections, Hitchin systems, integrable systems, Teichm\"uller theory,..., the list goes on. Third, and perhaps most importantly,
we have seen that the mere existence of the six-dimensional $(2,0)$
 theories leads to a host of nontrivial results in quantum field theory.
 Indeed, in this brief review we have not mentioned a large body of parallel
 beautiful and nontrivial  work on d=4 N=2 theories which has been done over
  the past few years by many  physicists. All this progress sharply intensifies
  the urgency of the open problem of formulating 6-dimensional superconformal theories in a mathematically
precise way. Many physicists regard this as one of the most
 outstanding problems in physical mathematics.

\section*{Acknowledgements}

The author heartily thanks Davide Gaiotto and Andy Neitzke for a very
productive collaboration leading to the papers
\cite{Gaiotto:2008cd,Gaiotto:2009hg,Gaiotto:2010be,Gaiotto:2011tf,Gaiotto:2012rg,Gaiotto:2012db}
  reviewed above.
He is also indebted to N. Seiberg and E. Witten for many explanations about N=2 theory.
He would also  like to thank D. L\"ust and I. Brunner for hospitality
at the Ludwig-Maximilians-Universit\"at M\"unchen,  where this talk was written.
This work is supported by the DOE under grant
DE-FG02-96ER40959.  The author also gratefully acknowledges hospitality of the
Institute for Advanced Study.  This work was partially supported by a grant
from the Simons Foundation (\#227381 to Gregory Moore).

\end{document}